\documentclass[preprint]{elsarticle}

\usepackage{gensymb}
\usepackage{multirow}

\usepackage{float}
\usepackage[latin1]{inputenc}   
\usepackage{amsmath} 
\usepackage{epstopdf}           %
\usepackage{flushend}           %
\usepackage{hyperref}           %

\usepackage[T1]{fontenc}
\usepackage{xcolor}

\usepackage{amssymb}

\usepackage{lineno}

\usepackage[figuresright]{rotating}
\usepackage{color,soul}

\usepackage{subfigure}

\begin{document}
\begin{frontmatter}

\title{3D Microstructural and Strain Evolution During the Early Stages of Tensile Deformation}

\author[First,Second]{A. Zelenika}
\ead{zeleni@dtu.dk}
\author[First]{C. Yildirim}
\author[First]{C. Detlefs}
\author[First]{R. Rodriguez-Lamas}
\author[Third]{F. B. Grumsen}
\author[Second]{H. F. Poulsen}
\author[Third]{G. Winther}
\corref{cor1}

\cortext[cor1]{Corresponding Author}

\address[First]{European Synchrotron Radiation Facility, 71 Avenue des Martyrs, CS40220, 38043 Grenoble Cedex 9, France.}
\address[Second]{Department of Physics, Technical University of Denmark, 2800 Kgs. Lyngby, Denmark.}
\address[Third]{Department of Civil and Mechanical Engineering, Technical University of Denmark, 2800 Kgs. Lyngby, Denmark.}
\date{\today}

\begin{abstract}

Dislocation patterning and self-organization during plastic deformation are associated with work hardening, but the exact mechanisms remain elusive. This is partly because studies of the structure and local strain during the initial stages of plastic deformation has been a challenge. Here we use Dark Field X-ray Microscopy to generate 3D maps
of embedded $350 \times 900 \times 72 \,\mu\mathrm{m}^3$ volumes within three pure Al single crystals, all oriented for double slip on the primary and conjugate slip systems. These were tensile deformed by 0.6\%, 1.7\% and 3.6\%, respectively. Orientation maps revealed the existence of two distinct types of planar dislocation boundaries both at 0.6\% and 1.7\% but no systematic patterning. At 3.6\%, these boundaries have evolved into a well-defined checkerboard pattern, characteristic of Geometrically Necessary Boundaries, GNBs. The GNB spacing is $\approx$ 14 $\mu$m and the misorientation $\approx$ 0.2$\degree$, in fair agreement with those at higher strains. By contrast to the sharp boundaries observed at higher strains, the boundaries are associated with a sinusoidal orientation gradient. Maps of the elastic strain along the (111) direction exhibit fluctuations of $\pm 0.0002 $ with an average domain size of 3 $\mu$m.

\end{abstract}


\end{frontmatter}

\section{Introduction}

The plastic deformation of a crystal lattice is caused by the motion of linear defects in the lattice, dislocations. These dislocations multiply and self-organize with increasing deformation into 3D structures of dislocation boundaries. In our current understanding, this is a multi-scale process leading to work hardening \cite{Hansen2011}. Well-established dislocation patterns at strains above about 5\% have been studied extensively, leading to classification of patterns according to crystallographic orientations \cite{huang2007,Le2012}. In addition, empirical rules for the dependence on slip systems \cite{Winther2007} have been established along with scaling laws \cite{Hughes1997, Godfrey2000} for the misorientation across boundaries and the spacing between boundaries. Furthermore, in this strain range the patterns exhibit self-similarity \cite{Hansen2011}. By contrast, there is a noticeable scarcity of data at lower strains where patterning is initiated. Provision of such data  may shed light on the underlying mechanisms.

Transmission Electron Microscopy (TEM) is the most widely-used technique to study dislocations, revealing both dislocation structures/walls (patterns) and isolated dislocations thanks to the high spatial resolution in combination with diffraction contrast \cite{hata2020electron}. One of the main downsides of TEM is the need for thin foils with a thickness of the order of 100\,nm or less. Such foils are not necessarily representative of the bulk, since the dislocations can interact with the free surface. In addition, the thin foil may not contain any boundaries if the boundary spacing is high and the boundaries are not steeply inclined to the plane of the foil. 

Other techniques such as Electron Backscatter Diffraction (EBSD) \cite{han2019ebsd}, Diffraction Contrast Tomography (DCT) \cite{stinville2022dct}, and 3D X-Ray Diffraction (3DXRD) \cite{pokharel2015hedm} have all provided valuable insights, particularly into stage III deformation on polycrystals. However, these techniques either lack the angular or the direct space resolution to study the patterning in the initial stages (i.e.~stage I and II) of deformation.

Recently, Dark Field X-ray Microscopy (DFXM), a diffraction-based imaging technique, has been used to image dislocations deeply embedded in mm-sized samples \cite{Jakobsen2019, Dresselhaus2021, yildirim2022extensive}. DFXM is analogous to dark-field TEM (i.e.~using an objective lens), but with a superior angular resolution ($\sim 10^{-3}$ $\degree$) that allows for accurate strain and orientation mapping of dislocations, while affording a spatial resolution on the order of 100\,nm \cite{Simons2015, Kutsal2019, yildirim2020probing}. Using an objective lens with a smaller numerical aperture than that of TEM, a single Bragg reflection is imaged at a time, and variations of components of the local strain and orientation around a given diffraction vector are mapped.

In this study, we investigate the unexplored: dislocation patterning in aluminum at strains below 4\%, using DFXM. We specifically focus on  well-annealed single crystals with a known orientation prior to deformation. We report on the initial patterning of dislocations and elastic strain after tensile loading.

\begin{figure}[h!]
    \centering
    \resizebox{1\columnwidth}{!}
    {\includegraphics{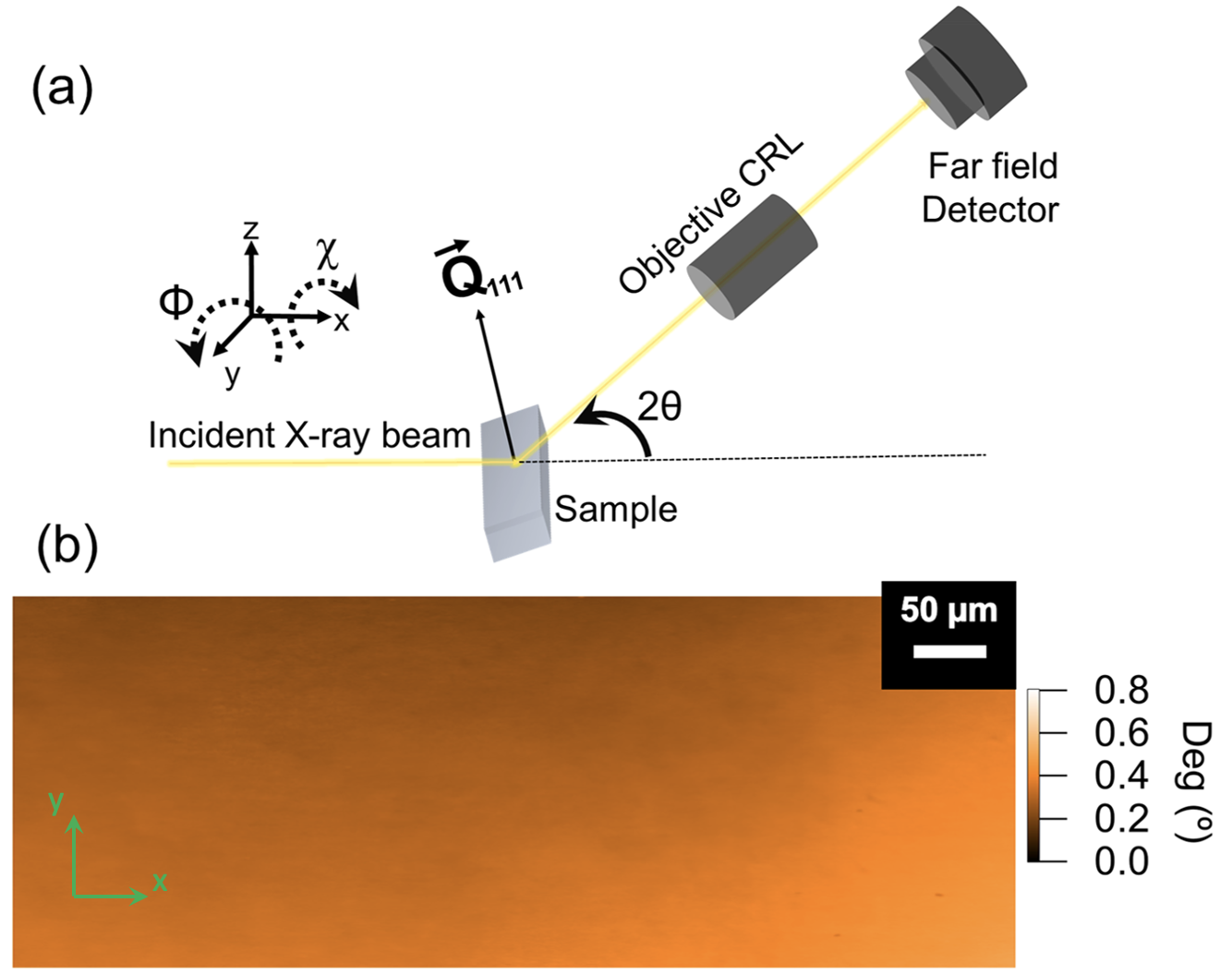}}
    \caption{(a) Schematic of the DFXM setup  with indication of the diffraction vector $\vec{Q}$, goniometer tilts $\phi,\chi$ and scattering angle $2\theta$. (b) $\chi$ center of mass (CoM) map of one layer in the undeformed sample. Before deformation no defects are visible.}
    \label{fig:setup}
\end{figure}

\section{Experimental Method}

Samples for tensile testing were cut out by Electrical Discharge Machining (EDM) from a 2.2\,mm thick sheet of AA1090 Aluminium exhibiting columnar grains with a diameter of several centimeters. Samples were cut out from a single grain with the tensile direction being close to $\langle 112 \rangle$ . The crystals are thus oriented for double slip on the primary and conjugate slip systems. Due to the shape of the grain, the tensile length varied between 35 and 50\,mm. After cutting, all samples were annealed at 540 $\degree$C for 10 hours. Tensile testing was performed to 0.6, 1.7 and 3.6 \% tensile elongation. Samples of mm$^3$ dimensions for DFXM were cut by EDM from the central part of the gauge section of these deformed specimens.

The DFXM experiments were conducted at Beamline ID06-HXM \cite{Kutsal2019, Poulsen2017} at the European Synchrotron Radiation Facility, ESRF. Figure~\ref{fig:setup}(a) shows a schematic of the experimental setup. We used a monochromatic beam with an energy of 17 keV. The beam was focused to a line with a FWHM of $\approx{}600$ nm in the $z$ direction, illuminating a layer within the sample. The 3D information is then acquired by translating the sample in the $z$ direction. For this study, we used the [111] Bragg reflection which is associated with a scattering angle of $2\theta=17.98\degree$. The objective was a Compound Refractive Lens (CRL), which magnified the diffracted signal by $\times 17.98$ onto a 2D detector located about 5 m from the sample. With an additional magnification of 2 in the camera itself the effective pixel size was  $\approx{}209 \times 679$ nm and the FOV in the sample is $350\,\mu\mathrm{m} \times 900\,\mu\mathrm{m}$. The details of the setup have been presented elsewhere \cite{yildirim2022extensive}.

To map the microstructure, we used two modes of DFXM: orientation and axial strain scans. Orientation scans comprise a mesh of the two orthogonal tilt angles of the sample (i.e.~$\phi$ and $\chi$, see Fig.\ref{fig:setup}(a)) to measure the variations of the local orientation around the (111) Bragg reflection. The strain scans involve a collective motion of the sample tilt $\phi$ and the $2\theta$ arm (i.e. objective and detector), mapping out the local difference in the d-spacing. Both orientation and strain scans were repeated for a series of z-positions.  The initial d-spacing was estimated as the mean value over the Field of View (FOV) assuming a macroscopically stress-free sample.

The orientation and strain maps that are used in this work are generated as center of mass (CoM) maps: given a series of images as function of say angle $\chi$, for each voxel one obtains a distribution of angles. The $\chi$ CoM map displays the average $\chi$ angle --- as such it is reminiscent of EBSD maps. All of the CoM analysis shown in this work is based on \textit{darfix}, a dedicated software package for DFXM data analysis \cite{darfix}. The correlation maps were analysed using MATLAB.

\section{Results}

We start by showing the microstructure of  an undeformed aluminium single crystal. Fig.~\ref{fig:setup}(b) shows the $\chi$ CoM map at $\phi$ and $2\theta$ fixed at the center position of the diffraction peak. Minor fluctuations in local orientation are clearly visible, as well as a small gradient going from top left to bottom right. By inspection of both the $\phi$ and $\chi$ maps we verified that this type of inhomogeneity is different from the deformation induced structure presented below and does not influence the conclusions drawn.

\subsection{Qualitative description of microstructure evolution}

\begin{figure}[!h]
    \centering
    \resizebox{1\columnwidth}{!}
    {\includegraphics{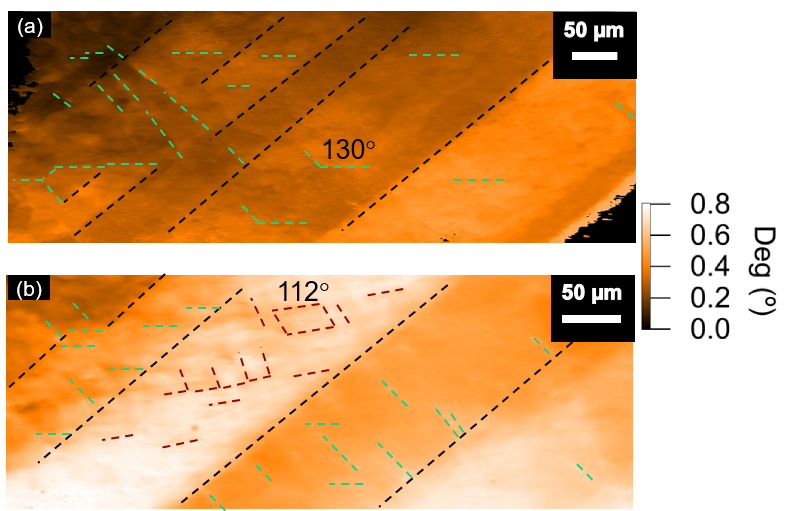}}
    \caption{$\chi$ CoM map of one layer in a) the 0.6\% deformed sample and (b) the 1.7\% deformed sample. Both a) and (b) have kink bands annotated with black dotted lines, and two sets of dislocation boundaries with 130$\degree$ between the traces annotated with dotted green lines. The red lines in (b) indicate dislocation boundaries with traces inclined 112$\degree$ to each other. The illuminated plane is inclined 9 $\degree$ to the [111] diffraction vector; normal to one of the active slip planes. The unannotated images are in the supplementary material.}
    \label{fig:small_def}
\end{figure}

In Fig.~\ref{fig:small_def}a) we show an orientation map for one layer in the 0.6\% deformed sample. The features marked by dashed black lines are nearly parallel. These primary features extend through the entire sample --- also in 3D. We also identify two types of more localized linear structures, annotated in green. We interpret these as two sets of emerging dislocation boundaries. The angle between the traces is about 130$\degree$. 

Likewise in Fig.~\ref{fig:small_def}(b) we show an orientation map for one layer in the 1.7\% deformed sample. The same primary and secondary structures are visible. 
A wide band with a different structure and delineated by the primary boundaries is visible in the left part of the image. In this band, the angle between the dislocation boundary traces  marked by red is smaller.

\begin{figure}[!h]
    \centering
    \resizebox{1\columnwidth}{!}
    {\includegraphics{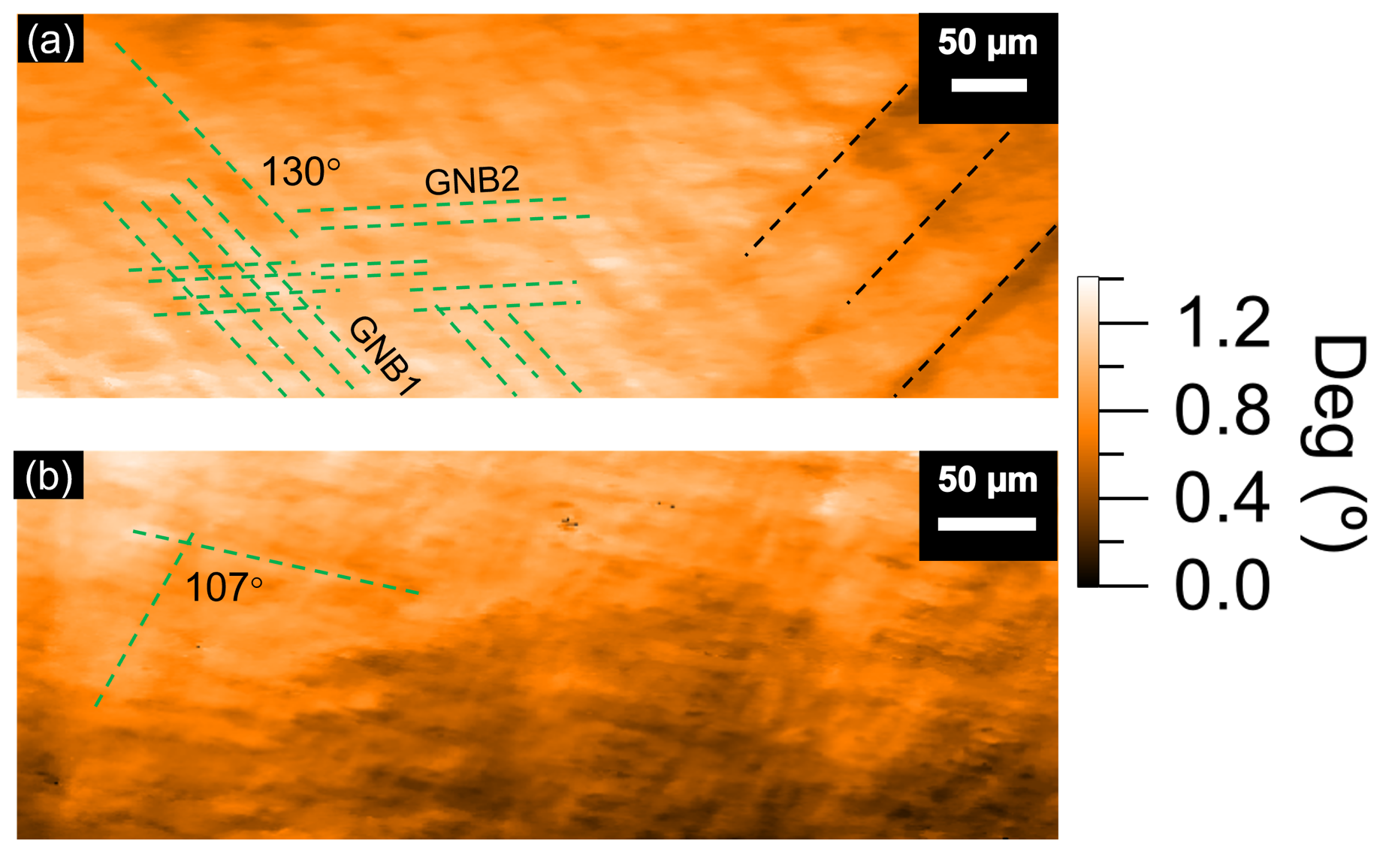}}
    \caption{Two $\chi$ CoM maps of two different \{111\} reflections of the  3.6\% deformed sample. The legend and annotations are similar to those of Fig.  \ref{fig:small_def}. }
    \label{fig:3}
\end{figure}

In Fig. \ref{fig:3}a) we show  corresponding orientation maps for one layer in the 3.6\% deformed sample. 
The dislocation structure has developed as compared to Fig.~\ref{fig:small_def}(b) into a checkerboard pattern. Alternating white and orange bands in between the boundaries marked by the green lines are now clearly visible, although the boundary between white and orange is not sharply defined. The alternating colours indicate alternating signs of the misorientation across these boundaries. At this strain, the original DFXM detector images also clearly reveal mis-orientation across the boundaries. This is in agreement with what is typically found for extended planar boundaries (termed geometrically necessary boundaries, GNBs) in electron microscopy at higher strains \cite{liu1998}. One of the two sets (labelled GNB1) is more pronounced than the one labelled GNB2. The spacing between the GNB1 set is higher.

Fig. \ref{fig:3}(b) is a similar plot acquired using a different \{111\} reflection of the same sample. Here the angle between the traces of the dislocation boundaries is smaller. As the two images in Fig. \ref{fig:3} do not correspond to the same position within the sample, it is possible that Fig. \ref{fig:3}(b) comes from a domain similar to the deformation band in Fig. \ref{fig:small_def}(b), which also had smaller angles between the boundary traces.  It is also possible that they correspond to the boundaries from Fig.  \ref{fig:3}a seen from another angle.  (The original CoM maps for both (a) and (b) with the full FOV and without annotation are available in the supplementary information.)

\begin{figure}[!h]
    \centering
    \resizebox{1\columnwidth}{!}
    {\includegraphics{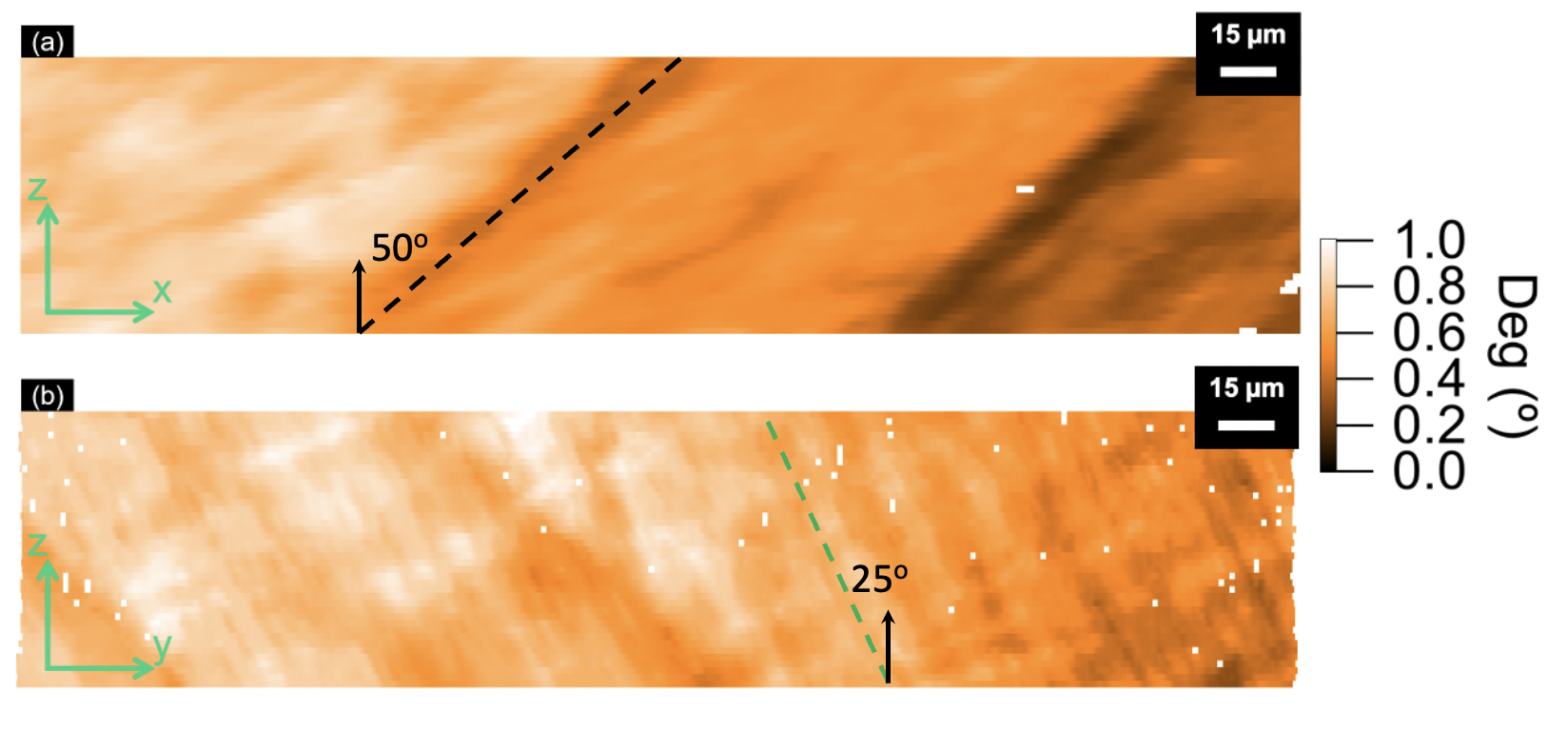}}
    \caption{Slices in the two directions orthogonal to Fig.~\ref{fig:3}(a) through the center of the 
   imaged 3D volume of the first reflection of the 
    3.6\% deformed sample. In (a) one kink band is annotated as a black dashed line while in (b) one dislocation
boundary is annotated as a dotted green line. The angle with the z-axis is 50 $\deg$ and 25 $\deg$, respectively. The z-axis is inclined 9 $\degree$ to the [111] diffraction vector. The color scale is identical to that of Fig.~\ref{fig:3}.}
    \label{fig:3Dmap}
\end{figure}

We emphasize that the DFXM maps are 3D. In 
Fig.~\ref{fig:3Dmap} we show results for a  $340 \times 340 \times 72 \mu$m$^3$ local volume of the 3.6 \% deformed sample. This represents the stacking of the $\chi$ COM maps for 41 layers in $z$ with an equidistant spacing of 1.8 $\mu$m. (The movement of the sample was not exactly along the lab z axis due to the rotation of the $\phi$ motor, so the images were shifted to account for a minor component of the  translation along the lab x axis.) It is clear from (a) that the kink bands extend through large volumes of the sample, at an angle of 50$\degree + \theta = 60\degree$  with the diffraction vector, $\vec{Q}_{111}$. Likewise from (b) it is evident that (at least) one set of GNBs extend over the entire z-range. Analysis of the GNB traces in the xy and zy planes confirms that GNB1 is roughly aligned with (11-1) and GNB2 with (111); the two active slip planes. 

\begin{figure}[!h]
    \centering
    \resizebox{1\columnwidth}{!}
    {\includegraphics{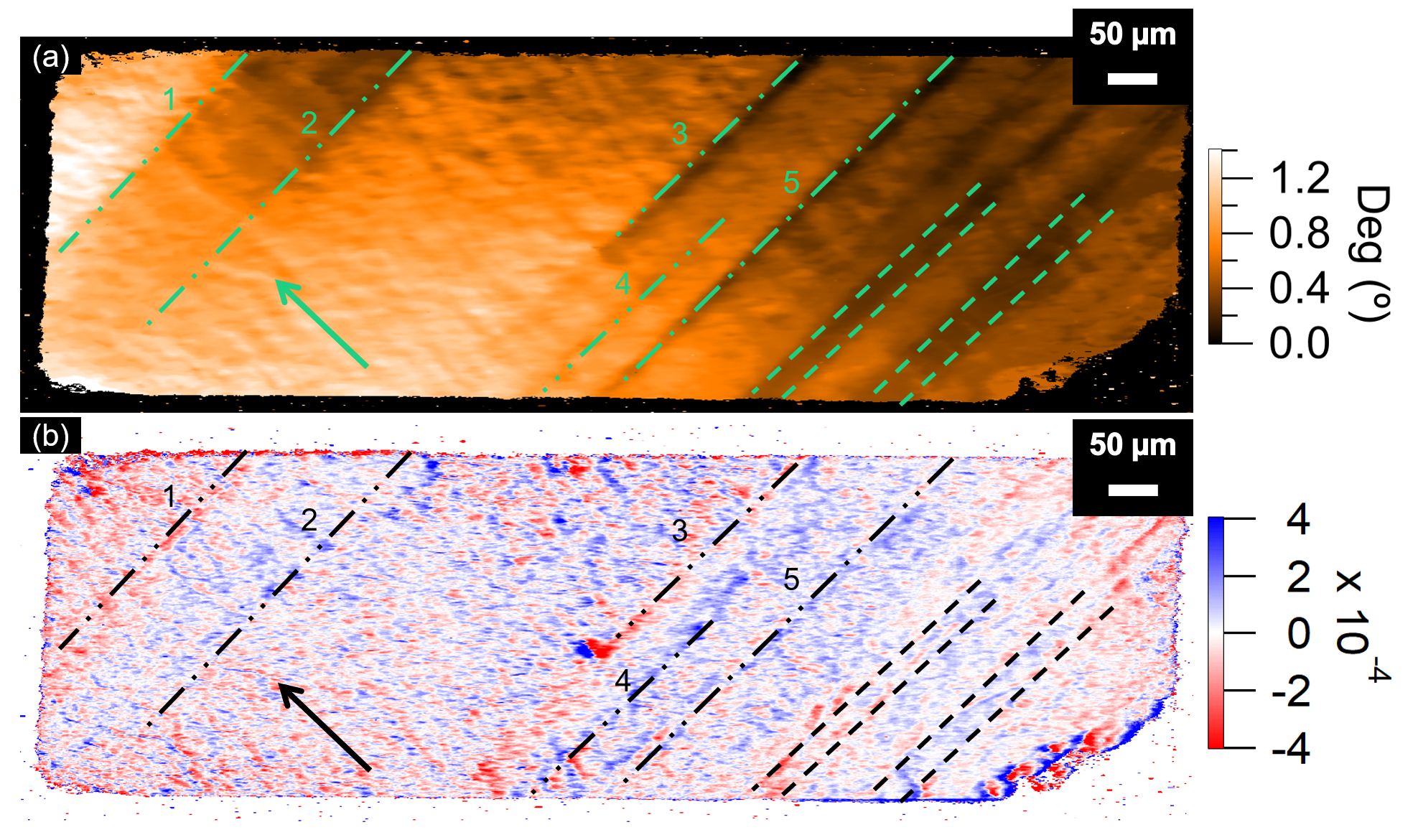}}
    \caption{(a) Orientation ($\chi$ CoM) map and (b) axial strain map of the entire Field-of-View for one layer of the 3.6 \% deformed sample (z=0). Units are degrees and strain, respectively. The primary bands defined by orientation changes and annotated by dashed lines in green in (a) are copied as dashed black lines in (b).  The arrow guides the eye to the directionality of the GNB1s, cf. Fig.~\ref{fig:3}(a) }
    \label{fig:strain_fov}
\end{figure}

\begin{figure}[!h]
    \centering
    \resizebox{1\columnwidth}{!}
    {\includegraphics{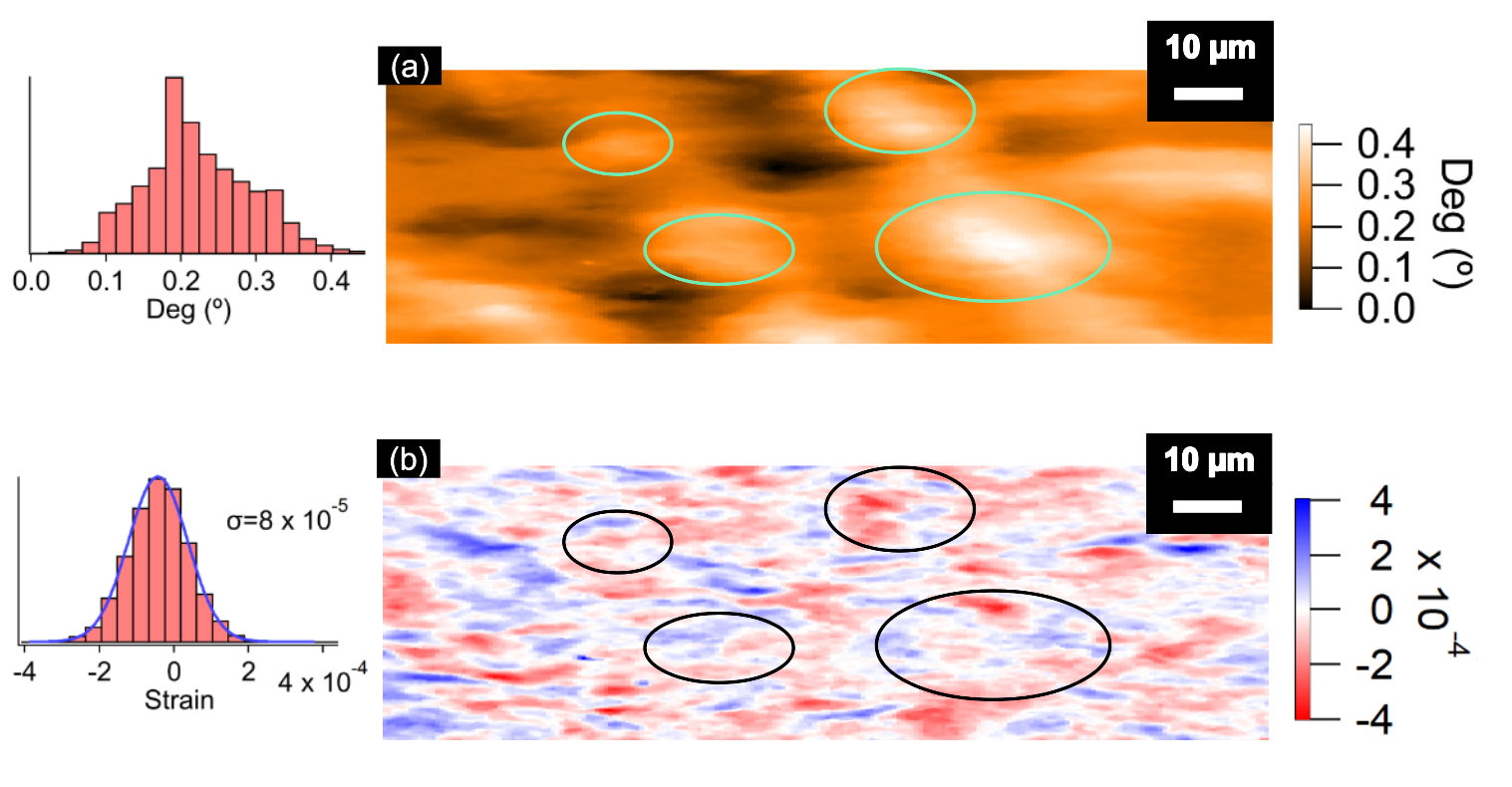}}
    \caption{Right: Comparison of maps of (a) local orientation ($\chi$ CoM map) and (b) elastic strain in direction of $\vec{Q}_{111}$ within a local region of the layer shown in Fig.~\ref{fig:strain_fov} for the 3.6\% deformed sample.  The colorscales are the same as used in Fig.\ref{fig:strain_fov}. The circles, which are placed in identical positions in (a) and (b) are guides to the eye only. Left: Statistics over microstructure for (a) and (b) respectively. (a) Histogram of $\chi$ values within the local Field-of-View. (b) Histogram of strain distribution and the corresponding best fit to a simple Gaussian distribution (blue line) with indication of the resulting standard deviation.}
    \label{fig:strain}
\end{figure}

As mentioned DFXM also readily provides a map of the axial (elastic) strain component in the direction of the diffraction vector, here $\vec{Q}_{111}$, by scanning $2\theta$. In Fig.~\ref{fig:strain_fov} we
compare the local variation in orientation (as determined by the CoM in the $\chi$-direction) and this axial strain in the entire Field-of-View of one layer of the 3.6 \% deformed sample. A local region of the orientation map for the same layer is shown in Fig.~\ref{fig:3} (a). Green lines in the orientation map and corresponding black lines in the strain map illustrate that the orientation change associated with the primary bands are also found in the strain map. Just like the orientations change at these bands, the sign of the strain in the immediate vicinity of them also alternates for those marked 1-3. Such a change is not clear for 4 and 5. Furthermore, for the unlabelled boundaries marked with only dashed lines in the right part of the images, the misorientation difference is clear whereas the general strain level is smaller and less correlated with the boundaries. Upon closer inspection of these boundaries in the detector images, the bands in this part have the character of a bundle of boundaries. In contrast with the GNB2s, the GNB1 directionality is also discernible in the strain map and marked by the arrow as a guide to the eye.

 Next, in Fig.~\ref{fig:strain} we zoom in on a region of interest in Fig.~\ref{fig:strain_fov}  that does not comprise the primary structure. By inspection we find that the domains in the orientation map are substantially larger than those in the strain map.  Notably, there is little visual correlation between the two local maps.   
 
 The structure displayed in Fig.~\ref{fig:strain} (a) and (b) are quantified in the left hand part of the figure. The elastic strain distribution is seen to be well described by a Gaussian with a $FWHM = 2.35 \sigma \approx 1.9\times 10^{-4}$.  The histogram for the $\chi$ distribution on the other hand exhibits a range that is 10 times larger, giving evidence of the evolution of orientation differences.

\subsection{Quantification by correlation functions}
 
\begin{figure}[!h]
    \centering
    \resizebox{1.0\columnwidth}{!}
    {\includegraphics{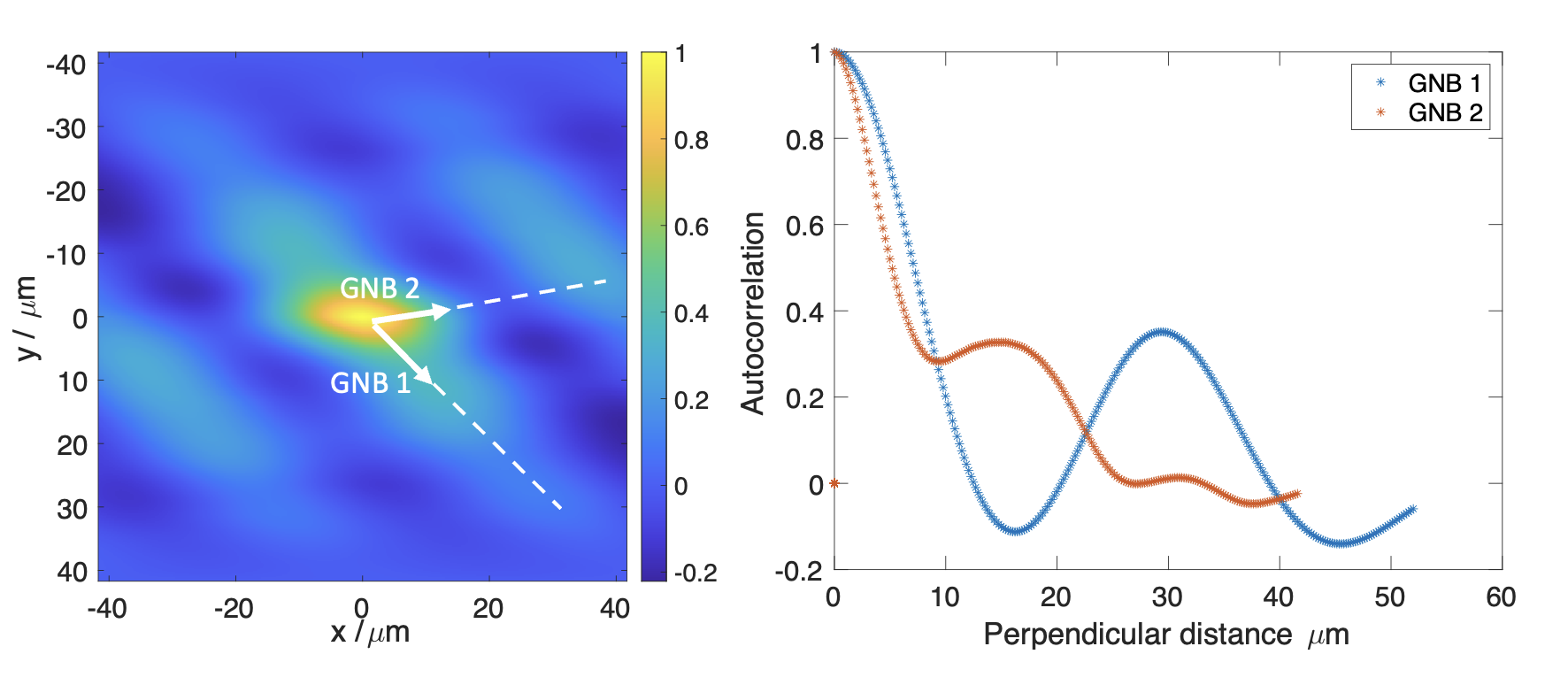}}
    \caption{Left: autocorrelation  of the orientation map displayed in Fig.~\ref{fig:strain} (a). Two main directions are identified in the pattern. Right: projections of this pattern perpendicular to the two directions marked by dashed lines.}
    \label{fig:corr_chi}
\end{figure}

In Fig.~\ref{fig:corr_chi}  the autocorrelation of the orientation map in the (x,y)-plane is shown for the 3.6 \% deformed sample (in the language of diffraction this is known as the point spread function). There is clear evidence for partial ordering in a structure with two basis vectors as marked in the figure. They exhibit an angle of 125$\degree$, in fair agreement with the roughly 130$\degree$ found above.  The GNB1 basis vector is rotated by 45$\degree$ with respect to the axes of the laboratory system. 

Next to the right in Fig.~\ref{fig:corr_chi} this 2D distribution is projected onto directions perpendicular to the two basis vectors. Both projections can be seen as modulated sinusoidal functions. The full period perpendicular to the GNB1 system is $\approx 29 \mu$m, and the corresponding domain size is therefore half of that: $\approx 14.5\mu$m. The ordering perpendicular to the GNB2  system is less prominent, but with a full period that within uncertainty is equal to half that of the full period perpendicular to the GNB1 system. 

The corresponding autocorrelation function of the orientation map in the (y,z)-plane is presented in the Supplementary Information. The results are consistent with extended planar boundaries that extend through the entire 3D region of interest.

We repeated the autocorrelation analysis for the samples elongated 0.6\% and 1.7\%. We find no signature for ordering in the orientation data of the kind shown in Fig.~\ref{fig:corr_chi}.

\begin{figure}[!h]
    \centering
    \resizebox{1.0\columnwidth}{!}
    {\includegraphics{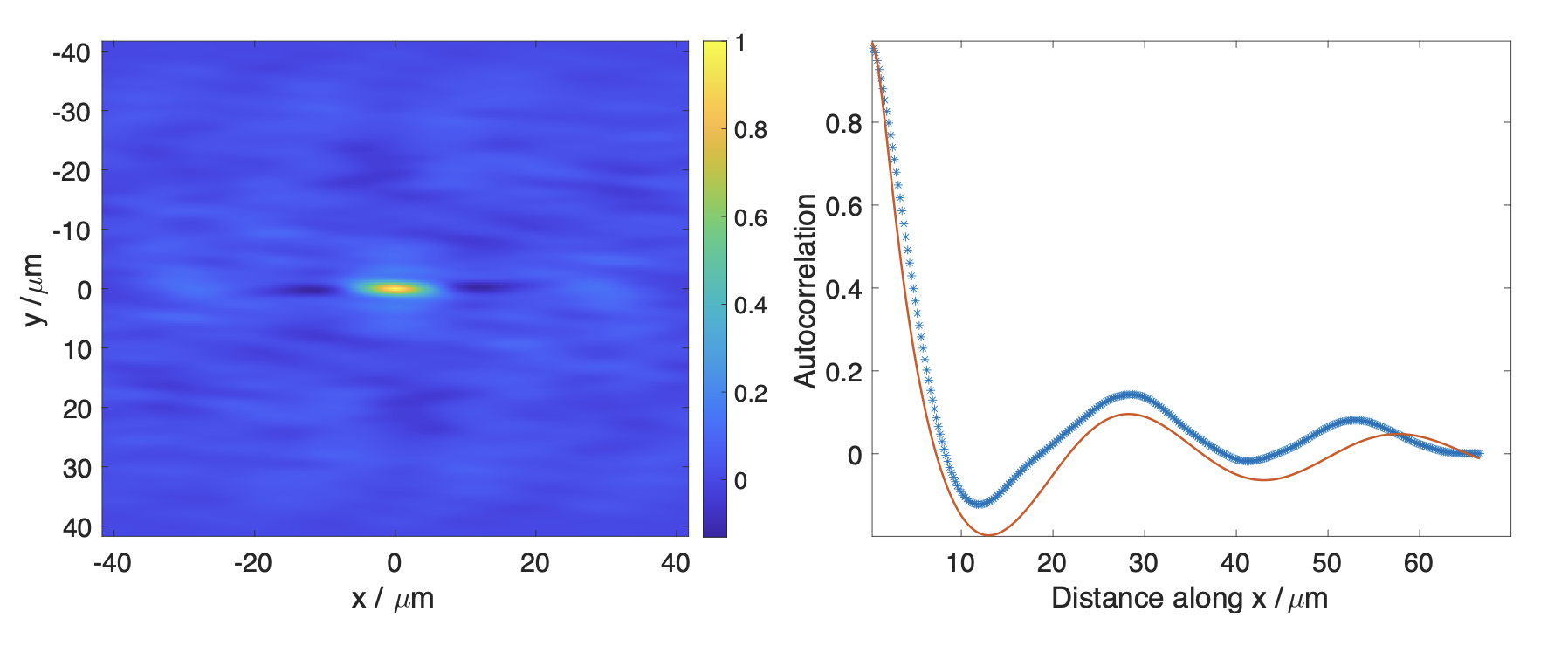}}
    \caption{Left: autocorrelation  of the strain map displayed in Fig.~\ref{fig:strain} (b). Right: The distribution for $y=0$ (blue dotes) along with a fit to a model of this distribution with a fixed period(red line) equal to the period of the blue curve in Fig.~\ref{fig:corr_chi}.}
    \label{fig:corr_strain}
\end{figure}

The corresponding autocorrelation function for the local strain map for the 3.6\% deformed sample, shown in Fig.~\ref{fig:strain} (b), is provided in Fig.~\ref{fig:corr_strain}. This appears less ordered except for a contribution directed along the sample x-axis. Again there is clear evidence of a modulated alternating plus/minus structure. As detailed in the supplementary information, by fitting this autocorrelation function the average size of the domains in the strain map is determined to be 3 $\mu$m, a factor of 5-10 smaller than the domains in the orientation map (Fig.~\ref{fig:strain} (a). 




\section{Discussion}

 \subsection{Instrumental}

This work demonstrates the relevance of using DFXM for studies of plastic deformation in the early strain regimes. The large volume inspected enable statistics and mapping on three length scales \emph{simultaneously}: the sample scale with an initially large spacing between the primary bands, the GNB structures and local variations within these. The high angular resolution allows the identification of structures at the very early stage and enable high accuracy quantitative measures for the local strain to be derived. The measurements are relatively fast, e.g. the data acquisition for the one-layer strain map shown in Fig.~\ref{fig:strain_fov} took 4 minutes. The application to \emph{in situ} studies is evident and has been pursued following this study. Correlating DFXM data directly to micro-mechanical models is discussed in \cite{Poulsen2021}.  

The main limitation of the technique in this study is the use of only one reflection for the same volume. This implies that only two out of three orientation variables and only one out of six strain components are probed.

 \subsection{Crystal subdivision}

 The large FOV combined with the high angular resolution reveals the subdivision of the crystal at two length scales. On the larger scale, primary boundaries with a spacing of the order of 50-200 $\mu$m are seen. The complex origin and nature of such widely spaced boundaries is treated in depth in \cite{Wert2005}. Notably, the present primary boundaries are neither classical kink bands perpendicular to the slip direction nor bands of secondary slip parallel to the slip plane. For the samples elongated by 0.6\%, 1.7\% and part of one at 3.6\%, these boundaries are associated with a misorientation of less than 0.1$\degree$ and the misorientation clearly alternates in sign across neighbouring boundaries. In some of these boundaries  individual dislocations could be discerned in the detector images. In another part of the sample at 3.6\% the misorientation is larger and these boundaries appear in bundles. It cannot be ruled out that this difference is a result of the larger strain.  However, we interpret them as boundaries formed in the initial stages of deformation. They probably form due to mechanical constraints from the grips and the differences between samples are likely to originate from small misalignment of the tensile axis in the testing machine.   
 These primary boundaries delineate bands within which subdivision takes place on a smaller length scale. As exemplified in Fig.~\ref{fig:small_def}b, the character of this subdivision is not always completely identical, giving evidence of deformation heterogeneity at the larger scale.

\subsection{GNBs}

The dislocation boundary structure on the smaller scale is clearly visible at 3.6\% strain with the characteristic parallel extended planar boundaries termed geometrically necessary boundaries that are also observed at higher strains. At 0.6\% and 1.7\%, the boundary structure is much less developed. It is, however, noteworthy that the same type of GNBs are visible. Still, distinct ordering has not yet taken place as evidenced by the autocorrelation analysis. 

Although the stacking fault energy of copper and aluminium is different, they have been found to exhibit the same alignment of GNBs. The directionality of GNBs in <112> oriented copper single crystals has been a topic of conflicting observations in previous studies using TEM. In one study [22], two sets of intersecting GNBs were observed, deviating 13$\degree$ from the slip planes, following straining to a stress of 74 MPa (estimated to correspond to approximately 10\% tensile elongation based on data from [23]). Interestingly, this deviation decreased with further straining. In contrast, another study [24] found GNBs aligned with the active slip planes at low strains. Similarly, variations in both deviation angle and direction have been reported in polycrystalline copper and aluminium, depending on the crystallographic direction of the tensile axis. Specifically, in polycrystals, the <112> orientation falls on the <001> - <111> line of the stereographic triangle, where a transition is observed from GNBs within 10$\degree$ of the slip planes to GNBs located further away (up to 29$\degree$) from the slip plane, accompanied by a change in deviation direction [2]. In this study, a consistent finding is the presence of boundaries deviating 5$\degree$-10$\degree$ from the slip plane traces for strains ranging from 0.6\% to 3.6\%. Notably, the direction of this deviation varies from domain to domain within the crystal, despite the domains having almost identical orientations. This discrepancy provides evidence of differences in slip system activity across the crystal. The observations presented here in aluminium contribute to a better understanding of the behavior of GNBs during plastic deformation and highlight the complexity of their response to strain in different crystallographic orientations. 

\begin{figure}[!h]
    \centering
    \resizebox{0.6\columnwidth}{!}
    {\includegraphics{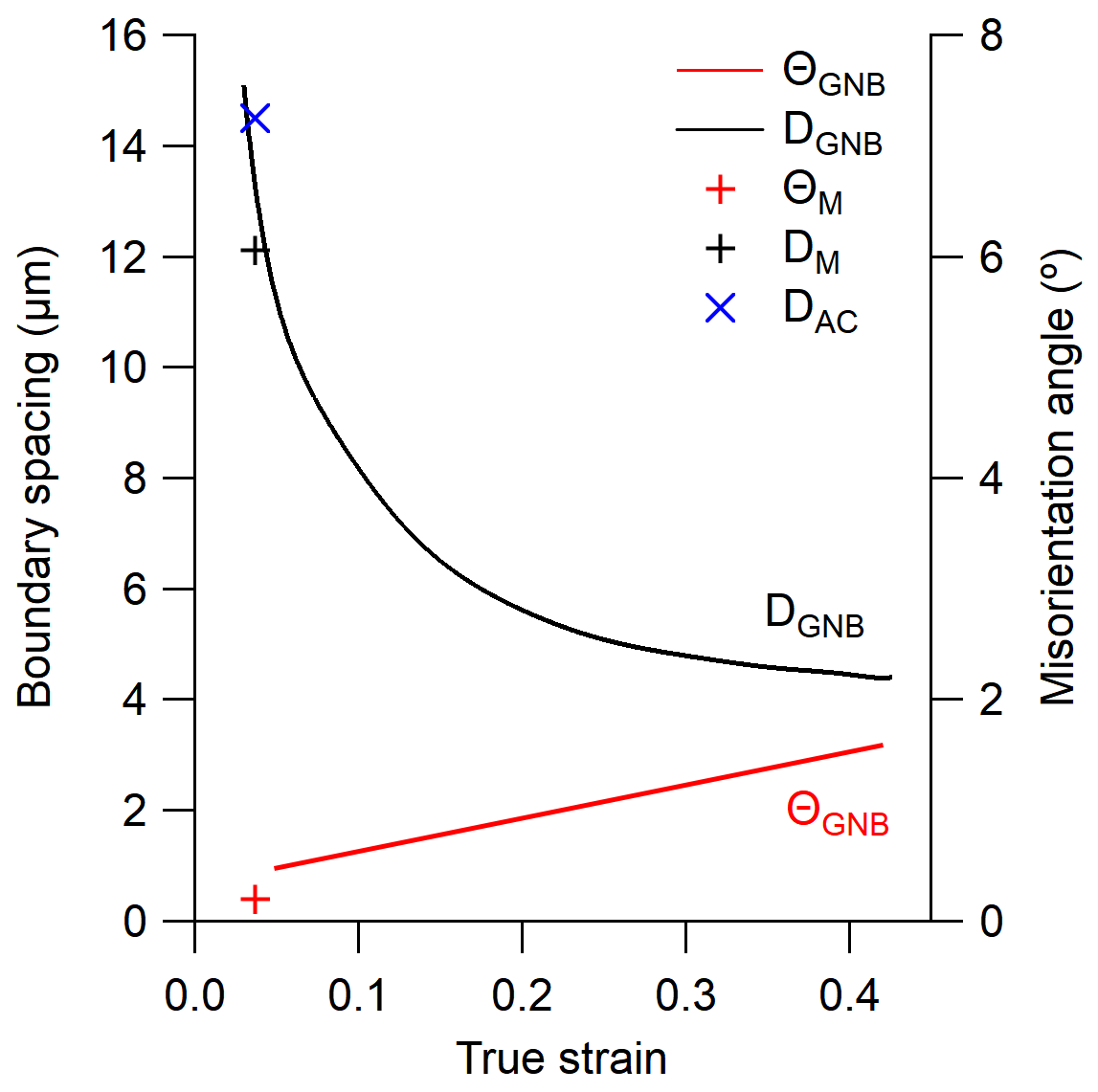}}
    \caption{Comparison of average GNB spacings and misorientations in the present experiment with literature data for 99.996\% pure tensile deformed aluminium polycrystals with GNBs of similar directionality \cite{Hansen2011}. The misorientation angles are denoted with the $\Theta$, while the boundary spacings are denoted with the D. The GNB subscript indicates the digitized data from the literature \cite{Hansen2011}, the M subscript indicates the data points obtained by the present study via manual inspection, and the AC subscript indicates the result obtained by autocorrelation.The errors of the two data points obtained by manual inspection have been determined as the standard deviation of the average, equaling 0.03\degree for the misorientation angle and 0.5 $\mu m$ for the boundary spacing.}
    \label{fig_scaling}
\end{figure}

With increasing strain, the dislocation density increases, which manifests itself in both smaller boundary spacings and higher misorientation angles. 
Fig. 9 shows that the spacing between GNB1s of half the period of 29 $\mu$m  for the 3.6\% elongated sample matches well with data in the literature for GNBs aligned within 10$\degree$ of slip planes in  99.996\% pure aluminium for tensile strains in the range 0.05-0.34 \cite{Hansen2011}. The mean misorientation angle of 0.21$\degree$ is slightly lower, which may be explained by the fact that the measured misorientation reflects only the spread of the probed 111 direction (and not rotations around it). (In a recent study it was shown that there is an underestimate of the misorientation measured by DFXM compared to EBSD measurements due to only a limited amount of the lattice curvature being measured in DFXM, and hence its magnitude is underestimated and so
is the geometrically necessary dislocation density.\cite{chen2023high}.)    

The good agreement with the quantitative data for 3.6\% strain and upwards and the similarity in directionality of the visually observed dislocation boundaries from 0.6\% to 3.6\% indicate that early patterning include mechanisms also active throughout the entire strain regime. Nevertheless, it is clear from the lack of order in autocorrelation for the 0.6 and 1.7\% elongated samples have not yet developed into regular patterns.

\subsection{Elastic strain level}

The maximum value of the elastic strain distribution in Fig.~\ref{fig:strain}(b) is $\approx 2\times 10^{-4}$. Using the diffraction elastic modulus of 72.8 GPa along the <111> direction \cite{zhang2020lattice}, this corresponds to a stress of 14 MPa. This is $\approx 70\%$ of the yield stress of 20 MPa. Stresses of 25\% the flow stress have been measured for a single reflection by a microbeam scanning technique in copper single crystals after 25\% elongation after unloading \cite{Levine2011}. Another study using the same technique but measuring three reflections found smaller stresses in the range 14-40\% of the flow stress in commercially pure aluminium grains subjected to equal-channel-angular pressing (ECAP) \cite{Phan2016}. The stress level decreased with the number of ECAP passes, i.e. plastic strain. The high stress level in the current study implies that mutual screening of the stress fields of the dislocations in the boundaries is limited. 

The mean misorientation angle in Fig.~\ref{fig:strain}(a) is 0.21\degree, which corresponds to a dislocation spacing in the boundaries of $\approx$ 80 nm. The dislocations in the network are not resolved and the full strain tensor is not probed. Nevertheless, comparison to the shear stress around a simple tilt boundary of edge dislocations with 80 nm spacing reveals that the stress more than 50 nm away from the boundary is below 14 MPa. Considering the spatial resolution of the strain data, this is in good agreement with the observations. 

Interestingly, the strain level obtained by Discrete Dislocation Dynamics (DDD) simulations of a copper single crystal tensile deformed to 0.85\% produced strains of $\approx 1.9\times 10^{-4}$ and misorientations of 0.01\degree \cite{Mohamed2015}. In the DDD simulation there was clear differences between the spatial distributions of the strain and orientation data. In the present experimental data, spatial correlation is clear for some of the primary boundaries whereas the orientation and strain differences manifest themselves with a common directionality in the large FOV in Fig.~\ref{fig:strain_fov}. In the zoomed-in images in \ref{fig:strain}, this visual correlation is lost. The correlation analysis of the zoomed-in strain data found a periodicity at a length scale that roughly matches that of the orientation data. However, for the strain the correlation is along a different direction. This is attributed to the complex strain fields around individual dislocations. The different orientation of GNB1 and GNB2 with respect to the diffraction vector probably explains why visual correlation was clearer for GNB1. 
The finding of short-range ordering of the stress with a domain size of 3 $\mu$m is evidence of the presence of additional dislocations in between the GNBs. 

\subsection{Sinusoidal shape of correlation functions}

Although the GNBs are clearly visible in the CoM map of the 3.6\% elongated sample, the sinusoidal shape of the autocorrelation function of this map reveals that these are not strictly two-dimensional but consist of a distribution of dislocations around a plane. This is in contrast to GNBs in aluminium at higher strains. The internal dislocation network within  GNBs in 10\% rolled aluminium has been observed to consist of a fairly regular grid of straight dislocation lines forming an almost perfect planar boundary \cite{Hong2013}, which further fulfill the so-called Frank equation for boundaries free of long-range stresses \cite{Winther2015}. By contrast, boundaries with a certain width are well-known from e.g. nickel, where boundary widths in the range 0.15-0.45 $\mu$m were measured in a 10\% cold rolled sample \cite{Hughes2000}. The boundary width rapidly decreased with further straining.  

The finding that the period of the orientation correlation is approximately twice that of the strain is in good agreement with the observation of alternating misorientation angles across the boundaries. This implies opposite signs of the dislocations in adjacent boundaries, meaning that shear strain fields of the same sign will extend into the domain between the boundaries.

\section {Conclusion}

The very onset of dislocation patterning has been investigated by DFXM in a tensile deformed aluminium single crystal oriented for double slip. The results reveal that GNBs are scarce but clearly visible already at 0.6\% and 1.7\% strain but with no systematic ordering into patterns. A well-ordered GNB pattern consisting of two sets of intersecting boundaries with alternating misorientation comparable to that at higher strains has formed at 3.6\% strain. The GNBs at 3.6\% strain have the same directionality as those at 0.6\% and 1.7\%. The misorientation and GNB spacing at 3.6\% are in good agreement with literature data for tensile deformed 99.996\% pure aluminium with boundaries close to slip planes. Yet, the boundaries are much more diffuse than those observed by TEM at higher deformation. 

The present findings reveal that patterning starts at the early stages of plastic deformation. Although the early patterns are not well-defined, they share characteristics with those at later stages, indicating that some of the fundamental mechanisms underlying patterning are the same.   

The axial strain along the probed crystallographic direction reveals some correlation with the primary bands and one set of GNBs in orientation data by visual inspection of a large domain. However, quantitative correlation on a local scale found ordering along a different direction. This is attributed to the complex character of the strain fields around dislocations. 

\section*{Acknowledgement}
The authors acknowledge the ESRF for provision of beamtime at ID06-HXM at the European Synchrotron Radiation Facility (ESRF). AZ, FBG, GW and HFP acknowledge support from ERC Advanced Grant nr 885022, from the ESS lighthouse on hard materials in 3D, SOLID, funded by the Danish Agency for Science and Higher Education, grant number 8144-00002B, and from DanScatt for travel cost.

\bibliography{references}

\end{document}